\documentclass[prb,twocolumn,showpacs,superscriptaddress,preprintnumbers,amssymb]{revtex4}
\usepackage{graphicx}
\usepackage{dcolumn}
\usepackage{bm}

\newcommand{\beq}{\begin{equation}}
\newcommand{\eeq}{\end{equation}}
\newcommand{\beqn}{\begin{eqnarray}}
\newcommand{\eeqn}{\end{eqnarray}}

\def\kappaETfull{{$\kappa$-(BEDT-TTF)$_{2}$Cu$_{2}$(CN)$_{3}$}}
\def\dmitfull{{EtMe$_{3}$Sb[Pd(dmit)$_{2}$]$_{2}$}}

\begin{document}

\title{Three Dimensional $Z_2$ Topological Phases enriched by Time-Reversal Symmetry }

\author{Cenke Xu}

\affiliation{Department of physics, University of California,
Santa Barbara, CA 93106, USA}

\begin{abstract}

We discuss three dimensional $Z_2$ topological phases enriched by
the $Z_2^T$ time-reversal symmetry with {\it bosonic} bulk
excitations. Some of these phases can be constructed by coupling
the three dimensional symmetry protected topological phases with
$Z_2 \times Z_2^T$ symmetry to a deconfined dynamical $Z_2$ gauge
field. Besides these simple phases, we also construct a special
root phase (phase {\bf 3}) which cannot be interpreted as gauged
$Z_2 \times Z_2^T$ SPT phase. The $2d$ boundary of these phases
can have an extra $Z_2$ topological order in addition to their
bulk topological order. In one of the possible boundary states of
this special root phase (phase {\bf 3}), $Z_2^T$ acts as the
``self-dual" transformation of the boundary topological order, in
the sense that the boundary $e$ and $m$ anyons are interchanged
under $Z_2^T$.

\end{abstract}

\date{\today}

\maketitle

\section{Introduction}

The interplay between symmetry and topology can lead to very rich
and profound physics. The most famous examples include the 2d
quantum spin Hall (QSH) insulator~\cite{kane2005a,kane2005b}, and
3d topological insulator
(TI)~\cite{moorebalents2007,roy2007,fukane}, where the charge
conservation and time-reversal symmetry together guarantee the
nontrivial spectrum at the boundary of the system. In the last few
years, QSH and TI have been generalized to strongly interacting
systems, and in general this type of phases were called symmetry
protected topological (SPT) phases. A SPT phase must have
completely trivial bulk spectrum, but nontrivial (either gapless
or degenerate) boundary spectrum when the system has certain
symmetry. In the last few years, SPT phase has rapidly developed
into a very active and exciting
field~\cite{wenspt,wenspt2,levingu,levinsenthil,levinstern,liuwen,luashvin,senthilashvin,xu3dspt,xu2dspt,xusenthil,wangsenthil,wangpottersenthil,chenluashvin,maxfisher,yewen1,yewen2,xuclass}.
So far we have understood many SPT phases, including their
mathematical classification~\cite{wenspt,wenspt2}, effective field
theories and bulk ground state wave
functions~\cite{senthilashvin,xu3dspt,xu2dspt,xusenthil,xuclass},
as well as the properties at the
boundary~\cite{senthilashvin,xuclass}.

SPT phases are short-range entangled states without intrinsic
topological order in the bulk. Recently a closely related concept
called symmetry enriched topological (SET) phase has also emerged
and soon attracted a lot of attentions. Unlike SPT phases, SET
phases have intrinsic topological orders in the bulk, and these
SET phases with the same bulk topological order are only distinct
from each other when the system has certain global symmetry $G_s$.
Mathematical classification of SET phases was studied in
Ref.~\onlinecite{hermeleset,ranset,wenset}, and according to
Ref.~\onlinecite{wenset}, SET phases with gauge group $G_g$ and
global symmetry $G_s$ are classified based on a larger group $G$
that satisfies $G/G_g = G_s$. However, physical properties of most
of these SET phases have not been explored yet.

In this work we study three dimensional $Z_2$ topological phases
with $Z_2^T$ time-reversal symmetry ($G_g = Z_2$, $G_s = Z_2^T$).
We expect these phases to be realizable in spin systems with local
Hamiltonians, and we will focus on the phases with bosonic bulk
excitations. The reason we choose to study SET phases with $G_g =
Z_2$ and $G_s = Z_2^T$ is the following:

{\it i.} $Z_2$ topological order is the simplest topological
order, and most likely to occur in nature. $Z_2$ topological order
have been confirmed numerically in various frustrated quantum spin
models and quantum boson
models~\cite{whitekagome,jiang1,jiang2,melko1,melko2}, and it was
proposed in many papers that the ground state of the organic
quantum frustrated spin systems
\kappaETfull~\cite{Shimuzi03_PRL_91_107001,Kurosaki05_PRL_95_177001}
and \dmitfull~\cite{131dmit1,131dmit2,131dmit3} could be a spin
liquid with $Z_2$ topological
order~\cite{Grover10_PRB_81_245121,qixu,rudro,kivelson,xudid}.
These spin liquid phases can be viewed as spin symmetry enriched
$Z_2$ topological phases. An exactly soluble model of this type of
$Z_2$ spin liquid with SO(3) spin symmetry is given in
Ref.~\onlinecite{yaoqi}.

{\it ii.} Time-reversal symmetry $Z_2^T$ is the most natural
symmetry of strongly correlated spin systems, especially for
systems with heavy elements. Due to spin-orbit coupling, the SU(2)
spin symmetry may be broken, but time-reversal symmetry is still
preserved. Thus a spin system in the absence of external magnetic
field and magnetic impurities should have at least the
time-reversal symmetry. Thus the phases with $G_s = Z_2^T$ and
$G_g = Z_2$ are very natural SET phases that will most likely
occur in nature.

{\it iii.} According to Ref.~\onlinecite{wenset}, SET phases with
gauge group $G_g$ and global symmetry $G_s$ are classified based
on an extension group $G$ that satisfies $G/G_g = G_s$. For
example when $G_s = G_g = Z_2$, $G$ can be either $Z_2 \times Z_2$
or $Z_4$. When $G_s = Z_2^T$, one obvious option of $G$ is $Z_2
\times Z_2^T$. But we will demonstrate that in three dimensional
space there are SET phases that cannot be classified using $G =
Z_2 \times Z_2^T$.


The classification of 2d SET phases with $Z_2$ topological order
has been studied in detail in Ref.~\onlinecite{luset} using the
Chern-Simons formalism. We will focus on three dimension in this
paper. The goal of this paper is to classify the 3d
$Z_2^T-$enriched $Z_2$ topological phases with bosonic bulk
excitations (the cases with fermionic bulk excitations will be
briefly discussed in the end), and explore their properties using
their bulk field theories. In total we obtain three different root
phases, based on which other phases can be constructed. Please
note that one can construct several ``trivial" SET phases by
making a product between the ordinary $Z_2$ topological order and
a $Z_2^T$ SPT phase. But these phases are not considered as root
phases here because they do not introduce new physics beyond the
ordinary $Z_2$ topological order and SPT phases.

\begin{center}
\begin{tabular}{ | r | r | r | r | r |  }
  \hline
    & $(\mathcal{T}_e)^4$ & $(\mathcal{T}_m)^4$ & $(G_g^e)^2$ & $(G_g^m)^2$ \\
  \hline
  Phase 1 & $+1$ & $+1$ & $-1$ & $+1$  \\
  \hline
  Phase 2 & $+1$ & $+1$ & $-1$ & $-1$ \\
  \hline
  Phase 3 & $-1$ & $-1$ & $-1$ & $-1$
  \\
  \hline
\end{tabular}

\end{center}

Just like SPT phases, SET phases can have nontrivial boundary
states. The boundary of the root phases constructed in this paper
can have an extra $Z_2$ topological order in addition to the bulk
$Z_2$ topological order, and the boundary anyons will have
nontrivial transformations under both $Z_2^T$ and bulk $Z_2$ gauge
symmetry. The transformations of the boundary $e$ and $m$
excitations of these three root phases are summarized in the table
above, where $\mathcal{T}_{e}$, $\mathcal{T}_{m}$ and $G_g^e$,
$G_g^m$ denote the time-reversal and {\it bulk} $Z_2$ gauge
transformations on $e$ and $m$ boundary anyons respectively.

\section{Basic formalism}

Some of the 3d $Z_2^T-$enriched $Z_2$ topological phases are very
easy to construct: We can take the 3d symmetry protected
topological (SPT) phases with $Z_2 \times Z_2^T$ global symmetry,
and couple the lattice degrees of freedom that transform
nontrivially under $Z_2$ symmetry to a dynamical $Z_2$ lattice
gauge field, and the deconfined phase of the $Z_2$ gauge field is
the desired $Z_2^T-$enriched topological phase. SPT phases have
been systematically classified using semiclassical nonlinear sigma
models~\cite{xuclass}. 3d SPT phases with $Z_2 \times Z_2^T$
symmetry have $(\mathbb{Z}_2)^3$ classification, namely all the
phases can be constructed with three basic root phases, which can
be described by the disordered phase (the phase where the coupling
constant $g$ flows to infinity under renormalization group) of the
following O(5) nonlinear sigma model (NLSM) field
theory~\cite{xuclass}: \beqn \mathcal{S} = \int d^3x d\tau \
\frac{1}{g} (\partial_\mu \vec{n})^2 + \frac{i 2\pi}{ \Omega_4 }
\epsilon_{abcde} n^a
\partial_x n^b \partial_y n^c
\partial_z n^d \partial_\tau n^e, \label{o5nlsm} \eeqn
where $\vec{n}$ is a five component vector with unit length, and
$\Omega_4$ is the volume of a four dimensional sphere with unit
radius. These three root phases are described by the same field
theory Eq.~\ref{o5nlsm}, but different transformations of the
order parameter $\vec{n}$ under the $Z_2\times Z_2^T$ symmetry:
\beqn \mathrm{Phase} \ 0 &:& Z_2: \vec{n} \rightarrow \vec{n}, \ \
\ \mathcal{T}: \vec{n} \rightarrow - \vec{n}; \cr\cr
\mathrm{Phase} \ 1 &:& Z_2: n_1, n_2 \rightarrow - n_1,-n_2, \ \
n_a \rightarrow n_a (a = 3, 4, 5); \cr\cr && \mathcal{T} : \vec{n}
\rightarrow - \vec{n}; \cr\cr \mathrm{Phase} \ 2 &:& Z_2: n_a
\rightarrow - n_a (a = 1 - 4), \ \ n_5 \rightarrow n_5; \cr\cr &&
\mathcal{T} : n_{1,2} \rightarrow n_{1,2}, \ \ n_{3,4,5}
\rightarrow - n_{3,4,5}. \label{transformation} \eeqn These three
choices of transformations all keep the entire action
Eq.~\ref{o5nlsm} including the $\Theta-$term invariant under $Z_2
\times Z_2^T$, and these transformations lead to three different
root SPT phases, based on which all the seven nontrivial SPT
phases can be constructed. After coupling the components of
$\vec{n}$ that transform nontrivially under $Z_2$ to a dynamical
$Z_2$ gauge field (for example in phase$-1$, components $n_1$ and
$n_2$ are coupled to the $Z_2$ gauge field), we obtain three
different root phases for 3d $Z_2^T-$enriched $Z_2$ topological
phases. These are the phases that can be classified using group
cohomology of $G = Z_2 \times Z_2^T$. However, the order parameter
$\vec{n}$ in phase$-0$ does not couple to the $Z_2$ gauge field at
all, this means that it only gives us a phase with $Z_2 $ gauge
field and a SPT phase trivially weakly coupled together, which
does not contain any new physics. Thus we {\it only} count
$Z_2$-gauged phase$-1$ and $2$ as $Z_2^T$ enriched $Z_2$
topological phase, which leads to $(\mathbb{Z}_2)^2$
classification.

The phases mentioned above clearly do not exhaust all the possible
$Z_2^T-$enriched topological phases, because in Eq.~\ref{o5nlsm}
all components of $\vec{n}$ that couple to the $Z_2$ gauge field
carry an ordinary Ising representation of time-reversal symmetry
that satisfies $\mathcal{T}^2 = 1$, while in a $Z_2$ topological
phase, a $Z_2$ gauge charge can carry a projective (or fractional)
representation of $Z_2^T$ that satisfies $\mathcal{T}^2 = -1$,
thus a pair of $Z_2$ gauge charges together still carry an
ordinary Ising representation.

One standard way of introducing fractional degrees of freedom for
spin systems is the CP$^1$ formalism: \beqn \vec{n} = \frac{1}{2}
z^\dagger_{\alpha} \vec{\sigma}_{\alpha\beta} z_\beta,
\label{cp1}\eeqn where $\vec{n}$ is a unit-length three component
vector order parameter that changes sign under $Z_2^T$ (for
example $\vec{n}$ can be the ordinary N\'{e}el order parameter),
and $z_\alpha = (z_1, z_2)$ is a fractional bosonic field that
transforms under time-reversal $\mathcal{T}: z_\alpha \rightarrow
i\sigma^y_{\alpha\beta}z_\beta$, thus $\mathcal{T}^2 = - 1$.

The fractional field $z_\alpha$ is defined up to an arbitrary
local U(1) rotation: $z_\alpha \rightarrow e^{i\theta} z_\alpha$,
thus in principle $z_\alpha$ is effectively coupled to a U(1)
gauge field. However, the U(1) gauge field can be broken down to a
$Z_2$ gauge field by condensing a pair of $z_\alpha$ that is
invariant under time-reversal transformation, for example $
\epsilon_{\alpha\beta}\epsilon_{ij} z_{\alpha,i} z_{\beta,j}$,
where $i,j$ are two different sites. In this work we assume that
this condensation occurs at a high energy scale, thus at low
energy the system is described by $z_\alpha$ coupled to a
deconfined $Z_2$ gauge field. We can use the fractional degree of
freedom $z_\alpha$ to construct 3d NLSMs with a $\Theta-$term like
Eq.~\ref{o5nlsm} that is invariant under both time-reversal
transformation and $Z_2$ gauge symmetry. Since the $Z_2$ gauge
field is fully gapped, at the mean field level we can ignore the
gauge fluctuation, and just construct NLSMs that are invariant
under time-reversal and $Z_2$ {\it global} symmetry. These NLSMs
should correspond to the desired $Z_2^T-$enriched $Z_2$
topological phases.

\section{Phase 3}

All these phases that we consider in this paper have the same
field theory as Eq.~\ref{o5nlsm}, but $\vec{n}$ will transform
differently from Eq.~\ref{transformation}. In particular, the next
example we consider has the following transformations: \beqn &&
z_1 \sim n_1 + i n_2, \ \ z_2^\ast \sim n_3 + i n_4, \cr\cr
\mathcal{T} &:&  n_1 \rightarrow n_3, \ n_2 \rightarrow n_4, \ n_3
\rightarrow - n_1, \ n_4 \rightarrow - n_2, \cr\cr && n_5
\rightarrow - n_5. \label{transformation2} \eeqn Eq.~\ref{o5nlsm}
is invariant under these transformations. We will ignore the $Z_2$
gauge fluctuation tentatively, and the rest of the analysis is
similar to Ref.~\onlinecite{senthilashvin}. With $\Theta = 2\pi$
in the bulk theory Eq.~\ref{o5nlsm}, the 2+1d boundary of the
system is described by a 2+1d O(5) NLSM with a Wess-Zumino-Witten
(WZW) term at level $k = 1$: \beqn S &=& \int d^2x d\tau \
\frac{1}{g} (\partial_\mu \vec{n})^2 \cr\cr &+& \int_0^1 du \
\frac{i 2\pi}{ \Omega_4 } \epsilon_{abcde} n^a
\partial_x n^b \partial_y n^c
\partial_z n^d \partial_\tau n^e, \label{2do5nlsm} \eeqn where $\vec{n}(x, \tau,
u)$ satisfies $\vec{n}(x, \tau, 0) = (0,0,0,0,1)$ and $\vec{n}(x,
\tau, 1) = \vec{n}(x, \tau)$. If the time-reversal symmetry is
preserved, namely $\langle n_5 \rangle = 0$, we can integrate out
$n_5$, and Eq.~\ref{2do5nlsm} reduces to a 2+1d O(4) NLSM with
$\Theta = \pi$: \beqn S = \int d^2x d\tau \ \frac{1}{g}
(\partial_\mu \vec{n})^2 + \frac{i \pi}{ 12\pi^2 }
\epsilon_{abcd}\epsilon_{\mu\nu\rho} n^a
\partial_\mu n^b \partial_\nu n^c
\partial_\rho n^d. \label{2do4} \eeqn In Eq.~\ref{2do4} $\Theta =
\pi$ is protected by time-reversal symmetry.

The field theory Eq.~\ref{o5nlsm} involves a $\Theta-$term in the
bulk, which is defined in 3+1d space-time based on homotopy group
$\pi_4[S^4] = \mathbb{Z}$. Let us couple two copies of
Eq.~\ref{o5nlsm} with $\Theta = 2\pi$ through a coupling allowed
by symmetry and gauge symmetry: \beqn \mathcal{S} &=&
\mathcal{S}[\vec{n}(x,\tau)_1, \Theta = 2\pi] +
\mathcal{S}[\vec{n}(x,\tau)_2, \Theta = 2\pi] \cr\cr &+& \int d^3x
d\tau \ h \vec{n}_1 \cdot \vec{n}_2. \label{couple} \eeqn Then
when $h < 0$, $\vec{n}_1 \sim \vec{n}_2 = \vec{n}$, the two
$\Theta-$terms combine into one $\Theta-$term for $\vec{n}$ with
$\Theta = 4\pi$; while when $h >0$, $\vec{n}_1 \sim - \vec{n}_2 =
\vec{n}$, the two $\Theta-$terms effectively cancel each other,
and combine into a $\Theta-$term with $\Theta = 0$. Tuning $h$
between these two limits will not close the bulk gap. This
statement can be checked numerically without the topological term,
because a $\Theta-$term with $\Theta = 2\pi$ will not change the
bulk spectrum at all. Intuitively, Eq.~\ref{couple} without the
topological terms can be viewed as two ``layers" of classical $4d$
O(5) magnet with interlayer coupling $h$. If both layers are
strongly disordered, an interlayer coupling will not drive the
system into an ordered phase. When $h = 0$, the bulk theory
Eq.~\ref{couple} becomes two decoupled O(5) NLSMs with $\Theta =
2\pi$, and each decoupled NLSM remains gapped and nondegenerate in
its disordered phase. This implies that $\Theta = 4\pi$ and $0$
can be smoothly connected to each other without a bulk phase
transition, thus the root phase {\bf 3} contributes a
$\mathbb{Z}_2$ classification, despite the fact that the homotopy
group $\pi_4[S^4] = \mathbb{Z}$. This is similar to the case of
the 1d Haldane's phase, where a $\Theta-$term is defined based on
the fact $\pi_2[S^2] = \mathbb{Z}$, while the classification is
still $\mathbb{Z}_2$. More discussions about this fact can be
found in Ref.~\onlinecite{xuclass}.

\subsection{Possible boundary topological orders}

The topological terms in Eq.~\ref{2do5nlsm} and Eq.~\ref{2do4}
guarantee that the boundary cannot be gapped without degeneracy.
One particularly interesting possibility of the boundary is a
phase with 2d $Z_2$ topological order, which is on top of the bulk
$Z_2$ topological order~\cite{senthilashvin}. A 2d $Z_2$
topological phase has $e$ and $m$ excitations that have mutual
semion statistics~\cite{kitaev2003}. The semion statistics can be
directly read off from Eq.~\ref{2do4}: the $\Theta-$term in
Eq.~\ref{2do4} implies that a vortex of $(n_3, n_4)$ carries half
charge of $z_1$, while a vortex of $(n_1, n_2)$ carries half
charge of $z_2$, thus vortices of $z_1$ and $z_2$ are bosons with
mutual semion statistics. This statistics survives after $z_1$ and
$z_2$ are disordered at the boundary, then the disordered phase
must inherit the statistics and become a $Z_2$ topological
phase~\cite{senthilashvin}. The vortices of $(n_1, n_2)$ and
$(n_3, n_4)$ become the $e$ and $m$ excitations respectively.
Normally a vortex defect is discussed in systems with a U(1)
global symmetry. We do not assume such U(1) global symmetry in our
case, this symmetry reduction is unimportant in the $Z_2$
topological phase.

At the vortex core of $(n_3, n_4)$, namely the $m$ excitation,
Eq.~\ref{2do5nlsm} reduces to a $0+1d$ O(3) NLSM with a WZW term
at level 1~\cite{groversenthil}: \beqn \mathcal{S}_m = \int d\tau
\frac{1}{g} (\partial_\tau \vec{N})^2 + \int_0^1 du \frac{i2\pi
}{8\pi} \epsilon_{abc}\epsilon_{\mu\nu} N^a
\partial_\mu N^b
\partial_\nu N^c, \eeqn where $\vec{N} \sim (n_1, n_2, n_5)$. This
0+1d field theory describes a single particle moving on a 2d
sphere with a magnetic monopole at the origin. It is well known
that if there is a SO(3) symmetry for $\vec{N}$, then the ground
state of this 0d problem has two fold degeneracy, with two
orthogonal solutions \beqn && u_m = \cos(\theta/2)e^{i\phi/2}, \ \
\ v_m = \sin(\theta/2)e^{- i\phi/2}, \cr\cr && \vec{N} =
\left(\sin(\theta)\cos(\phi), \sin(\theta)\sin(\phi), \cos(\theta)
\right). \eeqn In our paper we always demand $\theta \in [0,
\pi]$. Likewise, the vortex of $(n_1, n_2)$ ($e$ excitation) also
carries a doublet $(u_e, v_e)$. Under the bulk $Z_2$ gauge
transformation, $\phi \rightarrow \phi + \pi$, thus $u_{e,m}$ and
$v_{e,m}$ carry charge $\pm 1/2$ of the bulk $Z_2$ gauge symmetry,
namely under the bulk $Z_2$ gauge transformation: \beqn Z_2:
U_{e,m} \rightarrow e^{i\sigma^z \pi/2} U_{e,m}, \eeqn where
$U_{e,m} = (u_{e,m}, v_{e,m})^t$. So the bulk $Z_2$ gauge symmetry
further fractionalizes at the boundary, and for both $e$ and $m$
anyons, the bulk $Z_2$ gauge transformation satisfies $(G_g)^2 =
-1$.

At the boundary of an ordinary bosonic SPT phase with $Z_2^T$
symmetry, the $e$ and $m$ excitations of the boundary $Z_2$
topological order are both Kramers' doublet~\cite{senthilashvin}.
In our case, Eq.~\ref{transformation2} implies that time-reversal
transformation $\mathcal{T}$ switches $e$ and $m$ excitations at
the boundary. More precisely, Eq.~\ref{transformation2} leads to
the following transformation: \beqn \mathcal{T}: U_e \rightarrow
\sigma^x U_m, \ \ \ U_m \rightarrow \sigma^y U_e, \eeqn namely one
of the possible boundary theories is a ``self-dual" $Z_2$
topological order, and the duality transformation between $e$ and
$m$ is the time-reversal transformation. Notice that when deriving
this transformation, we require $\theta_{e,m} \in [0, \pi]$. Then
under time-reversal $\theta_m \rightarrow \pi-\theta_e$, $\theta_e
\rightarrow \pi- \theta_m$, $\phi_m \rightarrow \phi_e + \pi$,
$\phi_e \rightarrow \phi_m$.

Both $U_e$ and $U_m$ excitations satisfy $\mathcal{T}^4 = \sigma^x
\sigma^y \sigma^x \sigma^y = - 1$, namely $Z_2^T$ symmetry further
fractionalizes at the boundary $Z_2$ topological order. The
condensation of either $e$ or $m$ anyon will automatically break
the time-reversal symmetry, thus the boundary cannot be driven
into a trivial confined nondegenerate state without breaking
time-reversal symmetry.

In the boundary $Z_2$ topological order, $e$ and $m$ excitations
are connected by the time-reversal transformation. Also, because
$e$ and $m$ are different topological sectors, no local
perturbation can mix $e$ and $m$. Thus as long as time-reversal
symmetry is preserved, $e$ and $m$ excitations must be degenerate
and orthogonal states.

The same bulk theory Eq.~\ref{o5nlsm} and
Eq.~\ref{transformation2} can lead to another different boundary
$Z_2$ topological order. For example, the $e$ and $m$ excitations
of the boundary $Z_2$ phase can correspond to vortices of $(n_1,
n_3)$ and $(n_2, n_4)$ respectively. Then under time-reversal
transformation, the $e$ and $m$ excitations transform as \beqn
\mathcal{T} : U_{e,m} \rightarrow \frac{1}{\sqrt{2}}(\sigma^x +
\sigma^y) U_{e,m}. \eeqn Both $e$ and $m$ excitations still
satisfy $\mathcal{T}^4 = -1$.


\subsection{Bulk ground state wave function}

An ordinary $Z_2$ topological order on a 3d lattice is described
by the following Hamiltonian: \beqn H = -
\sum_{\mathrm{plaquette}}\prod\sigma^z - \sum_{\mathrm{vertex}}
\prod\sigma^x. \label{z2gauge} \eeqn $\sigma^z$ and $\sigma^x$ are
Pauli matrices defined on the links of the lattice. The first term
of Eq.~\ref{z2gauge} is a sum of product of $\sigma^z$ around each
plaquettes of the lattice, while the second term is a sum of
product of $\sigma^x$ around each vertex. The ground state of the
first term of this Hamiltonian is highly degenerate: they are all
the configurations of $\sigma^z$ in which the product of
$\sigma^z$ is $+1$ on every plaquette. Obviously $\sigma^z = +1$
everywhere is one of the configurations. And all the other ground
state configurations of the first term can be obtained by flipping
$\sigma^z$ to $-1$ on a closed 2d membrane of the 3d lattice. The
second term will mix the configurations that minimize the first
term. Eventually the ground state wave function of this ordinary
$Z_2$ topological order is schematically \beqn |\Psi\rangle \sim
\sum | \mathrm{membrane} \rangle, \eeqn this is an equal weight
superposition of all the membrane configurations within the same
topological sector.

\begin{figure}
\includegraphics[width=2.5 in]{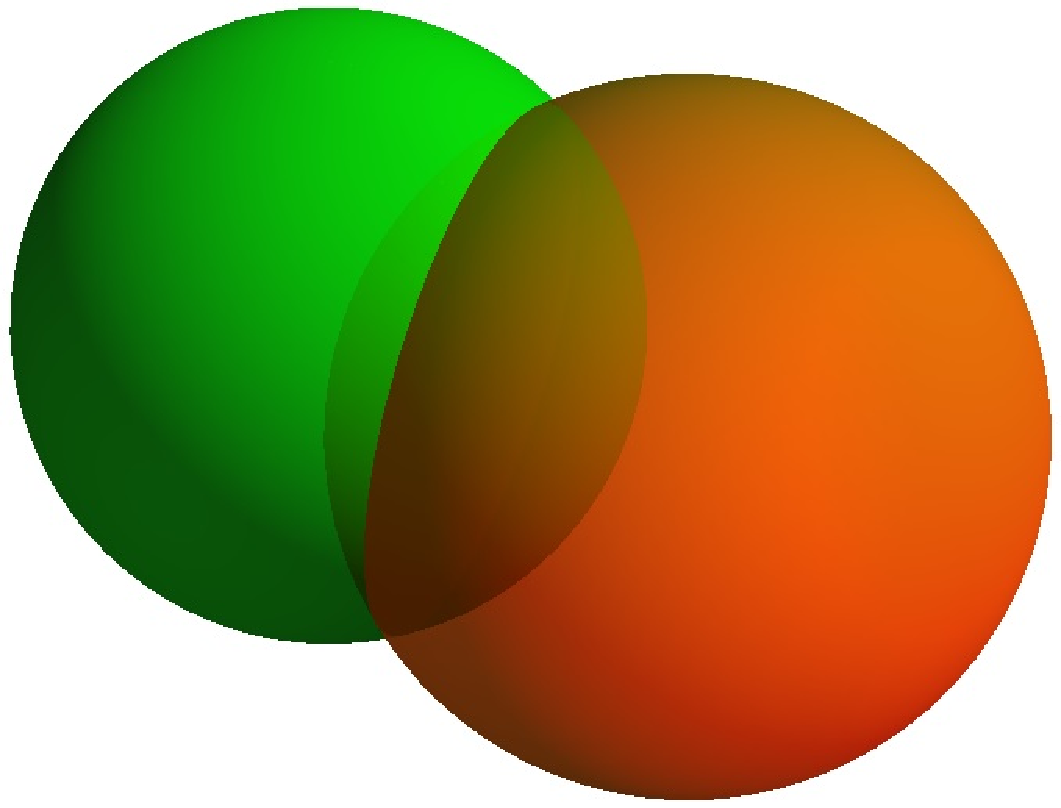}
\caption{The illustrative wave function of the SET phase discussed
in this section is a superposition of all the configurations of
$Z_2^T-$domain wall and $\sigma^z = -1$ membrane, with factor
$(-1)$ associated with each 1d intersection. Both domain wall and
membrane are 2d closed manifolds, they are illustrated by the two
spheres in this figure.} \label{membrane}
\end{figure}

The $Z_2^T-$enriched topological order is also a disordered phase
of time-reversal symmetry. The ground state wave function of an
ordinary time-reversal invariant phase can be naively written as
$|\Psi\rangle \sim \sum | \mathrm{domain \ wall} \rangle$, which
is a superposition of all the configurations of time-reversal
symmetry domain walls ($Z_2^T-$domain wall). We will argue that
the bulk ground state wave function of the phase discussed in this
section is \beqn |\Psi\rangle \sim \sum
(-1)^{\mathrm{intersection}} | \mathrm{membrane}, \ \mathrm{domain
\ wall} \rangle, \label{wave} \eeqn which is a superposition of
all the configurations of both $Z_2^T-$domain wall and $\sigma^z$
membranes, with a $(-1)$ coefficient contributed by each 1d
intersection between domain wall and membrane.

First of all, let us create a $Z_2^T-$domain wall, for example
$n_5$ has a domain wall in the XY plane $z = 0$: \beqn \vec{n}(x,
y, z = \pm \infty, \tau) = (0,0,0,0,\pm 1), \cr\cr \vec{n}(x, y, z
= 0, \tau) = (n_1,n_2,n_3,n_4,0); \eeqn then effectively the
theory that describes the $XY$ plane is a 2+1d O(4) NLSM for unit
vector $\vec{n} = (n_1, n_2, n_3, n_4)$ with $\Theta = 2\pi$. If
we ignore the bulk gauge fluctuation, $i.e.$ treat the bulk $Z_2$
gauge field at the mean field level, then the $Z_2$ gauge symmetry
becomes a $Z_2$ global symmetry of $\vec{n} = (n_1, n_2, n_3,
n_4)$. Because $n_a$ all change sign under this $Z_2$
transformation, this O(4) NLSM at the domain wall of $n_5$
describes a 2d SPT with $Z_2$ symmetry. The wave function of this
2d $Z_2$ SPT phase is $|\Psi\rangle \sim (-1)^{\# \mathrm{of} \
\mathrm{dw}} | Z_2 \ \mathrm{Config} \rangle$, which is a
superposition of all $Z_2$ Ising configurations, and dw denotes
the domain walls of the Ising
configurations~\cite{levingu,xusenthil}, thus each domain wall
will contribute factor $(-1)$ to the wave function. Now after we
couple the global $Z_2$ symmetry to a $Z_2$ gauge field, then a
$Z_2$ domain wall on a $Z_2^T-$domain wall is precisely a
intersection between the $Z_2^T-$domain wall and the membrane
$\sigma^z = -1$. Thus eventually we obtain the schematic wave
function Eq.~\ref{wave}.

\section{Two dimensional case}

The same strategy can be straightforwardly generalized to other
SET phases with different gauge and global symmetries, in all
dimensions. For example, let us consider two dimensional
$Z_2^T-$enriched $Z_2$ topological phases. The 2d SPT phases with
$Z_2 \times Z_2^T$ symmetry has a $(\mathbb{Z}_2)^2$
classification~\cite{wenspt,wenspt2}, and these phases can all be
constructed using the following two 2+1d O(4)
NLSMs~\cite{xuclass}: \beqn S &=& \int d^2x d\tau \ \frac{1}{g}
(\partial_\mu \vec{n})^2 + \frac{i 2 \pi}{ 12\pi^2 }
\epsilon_{abcd}\epsilon_{\mu\nu\rho} n^a
\partial_\mu n^b \partial_\nu n^c
\partial_\rho n^d, \cr\cr (1) &:& \ Z_2 : n_1, n_2 \rightarrow - n_1, -
n_2, \ \ n_3, n_4 \rightarrow n_3, n_4; \cr \cr && Z_2^T: n_1
\rightarrow n_1, \ \ n_a \rightarrow - n_a, (a = 3,4,5), \cr\cr
(2) &:& \ Z_2 : \vec{n} \rightarrow - \vec{n}, \cr \cr && Z_2^T :
n_1 \rightarrow n_1, \ \ n_a \rightarrow - n_a, (a = 3,4,5).
\label{2do42} \eeqn Two different 2d SET root phases can be
constructed by coupling the two SPT phases above to a deconfined
$Z_2$ gauge field.

In SET root phase (1), $e$ and $m$ anyons are $\mathcal{T}^2 = 1$
boson and $\mathcal{T}^2 = -1$ boson respectively. In this phase,
a vison is bound with half-vortex of $(n_1, n_2)$, and at the
half-vortex core of $(n_1, n_2)$, the $2+1d$ $\Theta-$term reduces
to a $0+1d$ $\Theta-$term with $\Theta_{0d} = \pi$: \beqn
\mathcal{S} &=& \int d\tau \ \frac{1}{g} (\partial_\tau \vec{n})^2
+ \frac{i\Theta_{0d}}{2\pi} \epsilon_{ab} n^a
\partial_\tau n^b, \ \ a, b = 3,4, \cr\cr \Theta_{0d} &=& \oint d\vec{l} \
\epsilon_{ef} n^e \partial_l n^f = \pi,  \ \ \ e,f = 1,2,
\label{thetap} \eeqn where $l$ is the line coordinate along a
large closed circle that links with the vison line. This $0+1d$
model can be solved exactly, and its ground state is two fold
degenerate, which is precisely a Kramers doublet.

SET root phase (2) is a double semion theory. In this phase, a
vison is bound with half-vortex of either $(n_1, n_2)$ or $(n_3,
n_4)$, and due to the existence of the $\Theta-$term, a vison will
carry a $Z_2$ gauge charge, which changes the statistics of the
vison to semion. Eq.~\ref{2do4} can be rewritten as a SU(2)
principle chiral model, by introducing SU(2) matrix field $G = n_4
\sigma^0 + i \vec{n} \cdot \vec{\sigma} $. $G$ has SU(2)$-$left
and SU(2)$-$right transformations: $G \rightarrow V^\dagger_L G
V_R$. Let us ``gauge" SU(2)$-$left and SU(2)$-$right
transformations with dynamical U(1) gauge fields $a_\mu \sigma^z$
and $b_\mu \sigma^z$, $i.e.$ replace $\partial_\mu G$ with
$\partial_\mu G + i a_\mu \sigma^z G + i b_\mu G \sigma^z$.
According to Ref.~\onlinecite{liuwen,levinsenthil}, after
integrating out matrix field $G$, gauge fields $a_\mu$ and $b_\mu$
both acquire a Chern-Simons term:  \beqn S_{cs} = \int d^2x d\tau
\ \frac{i2}{4\pi} \epsilon_{\mu\nu\rho} a_\mu
\partial_\nu a_{\rho} - \frac{i2}{4\pi} \epsilon_{\mu\nu\rho} b_\mu
\partial_\nu b_{\rho}. \label{cs}\eeqn This is because Eq.~\ref{2do4}
also describes a U(1) bosonic SPT phase with Hall conductivity 2.
The Chern-Simons action Eq.~\ref{cs} gives the unit gauge charges
of U(1) gauge field $a_\mu$ and $b_\mu$ a semion statistics, with
statistics angle $+\pi/2$ and $-\pi/2$ respectively. Notice that
the two U(1) gauge groups share the same $Z_2$ transformation $G
\rightarrow -G$, thus we can break the two U(1) gauge fields down
to one $Z_2$ gauge field, then the dynamical $\pi-$flux of this
$Z_2$ gauge field has two different flavors with semionic
statistics angle $+\pi/2$ and $-\pi/2$ respectively.

The same effect was discussed in Ref.~\onlinecite{levingu}, where
it was proved that after coupling the 2d SPT phase with $Z_2$
symmetry to a $Z_2$ gauge field, the system becomes a double
semion theory. In fact, the field theory of the SPT phase
discussed in Ref.~\onlinecite{levingu} is also an O(4) NLSM with
$\Theta = 2\pi$~\cite{xusenthil}.

Just like the 3d examples we discussed in previous section, we can
also try to construct 2d SET phases using fractionalized boson
$z_\alpha$ with $\mathcal{T}^2 = -1$. However, if we do not
distinguish between $e$ and $m$ anyons, then root phase (1) is
equivalent to the ordinary 2d $Z_2$ spin liquid with
$\mathcal{T}^2 = -1$ bosonic spinons $z_\alpha$, and it is the
only phase that we would be able to construct starting with
fractional boson $z_\alpha$, because starting with $z_\alpha$ one
cannot construct any 2+1d $\Theta-$term that is invariant under
$Z_2^T$. Thus NLSM constructed with $\mathcal{T}^2 = -1$
fractional bosons does not lead to any new $Z_2^T-$enriched $Z_2$
topological phase in 2d.

\section{Summary and Generalization}


So far we have constructed three different root phases. We can
weakly couple these root phases together and construct other
$Z_2^T-$enriched $Z_2$ topological phases. For example, phase$-1$
and $2$ in Eq~\ref{transformation} together give us four different
combinations and classification $(\mathbb{Z}_2)^2$, and phase-$3$
in section III and IV also give us classification $\mathbb{Z}_2$.
However, in one physical system, $Z_2$ gauge charges should have
either $\mathcal{T}^2 = 1$ or $-1$, but {\it not} both, because
two $Z_2$ gauge charges together is gauge neutral, thus they
should fuse into an ordinary representation of physical symmetry
$Z_2^T$. If the same $Z_2$ gauge field is coupled to two different
gauge charges with $\mathcal{T}^2 = 1$ and $-1$ respectively, then
two gauge charges can fuse into a gauge neutral excitation with
$\mathcal{T}^2 = -1$, which violates the basic rule. Thus
eventually the classification for $Z_2^T-$enriched $Z_2$
topological phases with bosonic bulk excitations is
$(\mathbb{Z}_2)^2 \oplus \mathbb{Z}_2$.

Throughout the paper we only considered SET phases with bosonic
bulk gauge charges. In a $Z_2$ topological order, the gauge
charges can also be fermions. However, the classification of such
phases requires a full understanding of strongly interacting SPT
phases of fermions with time-reversal symmetry. A fermionic gauge
charge can have either $\mathcal{T}^2 = +1$ or $-1$. For
$\mathcal{T}^2 = -1$ noninteracting fermions, there are infinite
number of 3d SPT phases (3d topological superconductors) with
classification
$\mathbb{Z}$~\cite{ludwigclass1,ludwigclass2,kitaevclass}. For
$\mathcal{T}^2 = +1$ fermions, there is no nontrivial 3d fermionic
SPT phase in the noninteracting case. Thus at the ``mean field
level", fermoinic gauge charges will contribute classification
$\mathbb{Z} \oplus \mathbb{Z}_1$, where $\mathbb{Z}$ comes from
coupling integer copies of topological superconductor
$\mathrm{He3B}$ to a deconfined $Z_2$ gauge field, and
$\mathbb{Z}_1$ comes from coupling the trivial
insulator/superconductor of $\mathcal{T}^2 = +1$ fermions to a
$Z_2$ gauge field. However, it is well-known that interacting
fermionic SPT phases can have a very different classification from
free fermions~\cite{fidkowski1,fidkowski2,qiz8,zhangz8}, and the
current understanding of 3d topological superconductors with
$\mathcal{T}^2 = -1$ is that, interaction reduces its
classification from $\mathbb{Z}$ to
$\mathbb{Z}_{16}$~\cite{chenhe3B,senthilhe3,youinversion,max14}.
But further study is required to finally determine whether there
is any strongly interacting fermionic SPT state without free
fermion analogue.

The author is supported by the the David and Lucile Packard
Foundation and NSF Grant No. DMR-1151208. The author also thanks
Meng Cheng, T. Senthil, Xiao-Gang Wen, and Ashvin Vishwanath for
very helpful discussions.

\bibliography{3dz2}

\begin{thebibliography}{56}
\expandafter\ifx\csname natexlab\endcsname\relax\def\natexlab#1{#1}\fi
\expandafter\ifx\csname bibnamefont\endcsname\relax
  \def\bibnamefont#1{#1}\fi
\expandafter\ifx\csname bibfnamefont\endcsname\relax
  \def\bibfnamefont#1{#1}\fi
\expandafter\ifx\csname citenamefont\endcsname\relax
  \def\citenamefont#1{#1}\fi
\expandafter\ifx\csname url\endcsname\relax
  \def\url#1{\texttt{#1}}\fi
\expandafter\ifx\csname urlprefix\endcsname\relax\def\urlprefix{URL }\fi
\providecommand{\bibinfo}[2]{#2}
\providecommand{\eprint}[2][]{\url{#2}}

\bibitem[{\citenamefont{Kane and Mele}(2005{\natexlab{a}})}]{kane2005a}
\bibinfo{author}{\bibfnamefont{C.~L.} \bibnamefont{Kane}} \bibnamefont{and}
  \bibinfo{author}{\bibfnamefont{E.~J.} \bibnamefont{Mele}},
  \bibinfo{journal}{Physical Review Letter} \textbf{\bibinfo{volume}{95}},
  \bibinfo{pages}{226801} (\bibinfo{year}{2005}{\natexlab{a}}).

\bibitem[{\citenamefont{Kane and Mele}(2005{\natexlab{b}})}]{kane2005b}
\bibinfo{author}{\bibfnamefont{C.~L.} \bibnamefont{Kane}} \bibnamefont{and}
  \bibinfo{author}{\bibfnamefont{E.~J.} \bibnamefont{Mele}},
  \bibinfo{journal}{Physical Review Letter} \textbf{\bibinfo{volume}{95}},
  \bibinfo{pages}{146802} (\bibinfo{year}{2005}{\natexlab{b}}).

\bibitem[{\citenamefont{Moore and Balents}(2007)}]{moorebalents2007}
\bibinfo{author}{\bibfnamefont{J.~E.} \bibnamefont{Moore}} \bibnamefont{and}
  \bibinfo{author}{\bibfnamefont{L.}~\bibnamefont{Balents}},
  \bibinfo{journal}{Physical Review B} \textbf{\bibinfo{volume}{75}},
  \bibinfo{pages}{121306(R)} (\bibinfo{year}{2007}).

\bibitem[{\citenamefont{Roy}(2009)}]{roy2007}
\bibinfo{author}{\bibfnamefont{R.}~\bibnamefont{Roy}},
  \bibinfo{journal}{Physical Review B} \textbf{\bibinfo{volume}{79}},
  \bibinfo{pages}{195322} (\bibinfo{year}{2009}).

\bibitem[{\citenamefont{Fu et~al.}(2008)\citenamefont{Fu, Kane, and
  Mele}}]{fukane}
\bibinfo{author}{\bibfnamefont{L.}~\bibnamefont{Fu}},
  \bibinfo{author}{\bibfnamefont{C.~L.} \bibnamefont{Kane}}, \bibnamefont{and}
  \bibinfo{author}{\bibfnamefont{E.~J.} \bibnamefont{Mele}},
  \bibinfo{journal}{Phys. Rev. Lett.} \textbf{\bibinfo{volume}{98}},
  \bibinfo{pages}{106803} (\bibinfo{year}{2008}).

\bibitem[{\citenamefont{Chen et~al.}(2013{\natexlab{a}})\citenamefont{Chen, Gu,
  Liu, and Wen}}]{wenspt}
\bibinfo{author}{\bibfnamefont{X.}~\bibnamefont{Chen}},
  \bibinfo{author}{\bibfnamefont{Z.-C.} \bibnamefont{Gu}},
  \bibinfo{author}{\bibfnamefont{Z.-X.} \bibnamefont{Liu}}, \bibnamefont{and}
  \bibinfo{author}{\bibfnamefont{X.-G.} \bibnamefont{Wen}},
  \bibinfo{journal}{Phys. Rev. B} \textbf{\bibinfo{volume}{87}},
  \bibinfo{pages}{155114} (\bibinfo{year}{2013}{\natexlab{a}}).

\bibitem[{\citenamefont{Chen et~al.}(2012)\citenamefont{Chen, Gu, Liu, and
  Wen}}]{wenspt2}
\bibinfo{author}{\bibfnamefont{X.}~\bibnamefont{Chen}},
  \bibinfo{author}{\bibfnamefont{Z.-C.} \bibnamefont{Gu}},
  \bibinfo{author}{\bibfnamefont{Z.-X.} \bibnamefont{Liu}}, \bibnamefont{and}
  \bibinfo{author}{\bibfnamefont{X.-G.} \bibnamefont{Wen}},
  \bibinfo{journal}{Science} \textbf{\bibinfo{volume}{338}},
  \bibinfo{pages}{1604} (\bibinfo{year}{2012}).

\bibitem[{\citenamefont{Levin and Gu}(2012)}]{levingu}
\bibinfo{author}{\bibfnamefont{M.}~\bibnamefont{Levin}} \bibnamefont{and}
  \bibinfo{author}{\bibfnamefont{Z.-C.} \bibnamefont{Gu}},
  \bibinfo{journal}{Phys. Rev. B} \textbf{\bibinfo{volume}{86}},
  \bibinfo{pages}{115109} (\bibinfo{year}{2012}).

\bibitem[{\citenamefont{Levin and Senthil}(2013)}]{levinsenthil}
\bibinfo{author}{\bibfnamefont{M.}~\bibnamefont{Levin}} \bibnamefont{and}
  \bibinfo{author}{\bibfnamefont{T.}~\bibnamefont{Senthil}},
  \bibinfo{journal}{Phys. Rev. Lett.} \textbf{\bibinfo{volume}{110}},
  \bibinfo{pages}{046801} (\bibinfo{year}{2013}).

\bibitem[{\citenamefont{Levin and Stern}(2012)}]{levinstern}
\bibinfo{author}{\bibfnamefont{M.}~\bibnamefont{Levin}} \bibnamefont{and}
  \bibinfo{author}{\bibfnamefont{A.}~\bibnamefont{Stern}},
  \bibinfo{journal}{Phys. Rev. B} \textbf{\bibinfo{volume}{86}},
  \bibinfo{pages}{115131} (\bibinfo{year}{2012}).

\bibitem[{\citenamefont{Liu and Wen}(2013)}]{liuwen}
\bibinfo{author}{\bibfnamefont{Z.-X.} \bibnamefont{Liu}} \bibnamefont{and}
  \bibinfo{author}{\bibfnamefont{X.-G.} \bibnamefont{Wen}},
  \bibinfo{journal}{Phys. Rev. Lett.} \textbf{\bibinfo{volume}{110}},
  \bibinfo{pages}{067205} (\bibinfo{year}{2013}).

\bibitem[{\citenamefont{Lu and Vishwanath}(2012)}]{luashvin}
\bibinfo{author}{\bibfnamefont{Y.-M.} \bibnamefont{Lu}} \bibnamefont{and}
  \bibinfo{author}{\bibfnamefont{A.}~\bibnamefont{Vishwanath}},
  \bibinfo{journal}{Phys. Rev. B} \textbf{\bibinfo{volume}{86}},
  \bibinfo{pages}{125119} (\bibinfo{year}{2012}).

\bibitem[{\citenamefont{Vishwanath and Senthil}(2013)}]{senthilashvin}
\bibinfo{author}{\bibfnamefont{A.}~\bibnamefont{Vishwanath}} \bibnamefont{and}
  \bibinfo{author}{\bibfnamefont{T.}~\bibnamefont{Senthil}},
  \bibinfo{journal}{Phys. Rev. X} \textbf{\bibinfo{volume}{3}},
  \bibinfo{pages}{011016} (\bibinfo{year}{2013}).

\bibitem[{\citenamefont{Xu}(2013)}]{xu3dspt}
\bibinfo{author}{\bibfnamefont{C.}~\bibnamefont{Xu}}, \bibinfo{journal}{Phys.
  Rev. B} \textbf{\bibinfo{volume}{87}}, \bibinfo{pages}{144421}
  (\bibinfo{year}{2013}).

\bibitem[{\citenamefont{Oon et~al.}(2012)\citenamefont{Oon, Cho, and
  Xu}}]{xu2dspt}
\bibinfo{author}{\bibfnamefont{J.}~\bibnamefont{Oon}},
  \bibinfo{author}{\bibfnamefont{G.~Y.} \bibnamefont{Cho}}, \bibnamefont{and}
  \bibinfo{author}{\bibfnamefont{C.}~\bibnamefont{Xu}},
  \bibinfo{journal}{arXiv:1212.1726}  (\bibinfo{year}{2012}).

\bibitem[{\citenamefont{Xu and Senthil}(2013)}]{xusenthil}
\bibinfo{author}{\bibfnamefont{C.}~\bibnamefont{Xu}} \bibnamefont{and}
  \bibinfo{author}{\bibfnamefont{T.}~\bibnamefont{Senthil}},
  \bibinfo{journal}{Phys. Rev. B} \textbf{\bibinfo{volume}{87}},
  \bibinfo{pages}{174412} (\bibinfo{year}{2013}).

\bibitem[{\citenamefont{Wang and Senthil}(2013)}]{wangsenthil}
\bibinfo{author}{\bibfnamefont{C.}~\bibnamefont{Wang}} \bibnamefont{and}
  \bibinfo{author}{\bibfnamefont{T.}~\bibnamefont{Senthil}},
  \bibinfo{journal}{Phys. Rev. B} \textbf{\bibinfo{volume}{87}},
  \bibinfo{pages}{235122} (\bibinfo{year}{2013}).

\bibitem[{\citenamefont{Wang et~al.}(2013)\citenamefont{Wang, Potter, and
  Senthil}}]{wangpottersenthil}
\bibinfo{author}{\bibfnamefont{C.}~\bibnamefont{Wang}},
  \bibinfo{author}{\bibfnamefont{A.~C.} \bibnamefont{Potter}},
  \bibnamefont{and} \bibinfo{author}{\bibfnamefont{T.}~\bibnamefont{Senthil}},
  \bibinfo{journal}{arXiv:1306.3238}  (\bibinfo{year}{2013}).

\bibitem[{\citenamefont{Chen et~al.}(2013{\natexlab{b}})\citenamefont{Chen, Lu,
  and Vishwanath}}]{chenluashvin}
\bibinfo{author}{\bibfnamefont{X.}~\bibnamefont{Chen}},
  \bibinfo{author}{\bibfnamefont{Y.-M.} \bibnamefont{Lu}}, \bibnamefont{and}
  \bibinfo{author}{\bibfnamefont{A.}~\bibnamefont{Vishwanath}},
  \bibinfo{journal}{arXiv:1303.4301}  (\bibinfo{year}{2013}{\natexlab{b}}).

\bibitem[{\citenamefont{Metlitski et~al.}(2013)\citenamefont{Metlitski, Kane,
  and Fisher}}]{maxfisher}
\bibinfo{author}{\bibfnamefont{M.~A.} \bibnamefont{Metlitski}},
  \bibinfo{author}{\bibfnamefont{C.~L.} \bibnamefont{Kane}}, \bibnamefont{and}
  \bibinfo{author}{\bibfnamefont{M.~P.~A.} \bibnamefont{Fisher}},
  \bibinfo{journal}{arXiv:1302.6535}  (\bibinfo{year}{2013}).

\bibitem[{\citenamefont{Ye and Wen}(2013{\natexlab{a}})}]{yewen1}
\bibinfo{author}{\bibfnamefont{P.}~\bibnamefont{Ye}} \bibnamefont{and}
  \bibinfo{author}{\bibfnamefont{X.-G.} \bibnamefont{Wen}},
  \bibinfo{journal}{Phys. Rev. B} \textbf{\bibinfo{volume}{87}},
  \bibinfo{pages}{195128} (\bibinfo{year}{2013}{\natexlab{a}}).

\bibitem[{\citenamefont{Ye and Wen}(2013{\natexlab{b}})}]{yewen2}
\bibinfo{author}{\bibfnamefont{P.}~\bibnamefont{Ye}} \bibnamefont{and}
  \bibinfo{author}{\bibfnamefont{X.-G.} \bibnamefont{Wen}},
  \bibinfo{journal}{arXiv:1303.3572}  (\bibinfo{year}{2013}{\natexlab{b}}).

\bibitem[{\citenamefont{Bi et~al.}(2013)\citenamefont{Bi, Rasmussen, and
  Xu}}]{xuclass}
\bibinfo{author}{\bibfnamefont{Z.}~\bibnamefont{Bi}},
  \bibinfo{author}{\bibfnamefont{A.}~\bibnamefont{Rasmussen}},
  \bibnamefont{and} \bibinfo{author}{\bibfnamefont{C.}~\bibnamefont{Xu}},
  \bibinfo{journal}{arXiv:1309.0515}  (\bibinfo{year}{2013}).

\bibitem[{\citenamefont{Essin and Hermele}(2013)}]{hermeleset}
\bibinfo{author}{\bibfnamefont{A.~M.} \bibnamefont{Essin}} \bibnamefont{and}
  \bibinfo{author}{\bibfnamefont{M.}~\bibnamefont{Hermele}},
  \bibinfo{journal}{Phys. Rev. B} \textbf{\bibinfo{volume}{87}},
  \bibinfo{pages}{104406} (\bibinfo{year}{2013}).

\bibitem[{\citenamefont{Mesaros and Ran}(2013)}]{ranset}
\bibinfo{author}{\bibfnamefont{A.}~\bibnamefont{Mesaros}} \bibnamefont{and}
  \bibinfo{author}{\bibfnamefont{Y.}~\bibnamefont{Ran}},
  \bibinfo{journal}{Phys. Rev. B} \textbf{\bibinfo{volume}{87}},
  \bibinfo{pages}{155115} (\bibinfo{year}{2013}).

\bibitem[{\citenamefont{Hung and Wen}(2013)}]{wenset}
\bibinfo{author}{\bibfnamefont{L.-Y.} \bibnamefont{Hung}} \bibnamefont{and}
  \bibinfo{author}{\bibfnamefont{X.-G.} \bibnamefont{Wen}},
  \bibinfo{journal}{Phys. Rev. B} \textbf{\bibinfo{volume}{87}},
  \bibinfo{pages}{165107} (\bibinfo{year}{2013}).

\bibitem[{\citenamefont{Yan et~al.}(2011)\citenamefont{Yan, Huse, and
  White}}]{whitekagome}
\bibinfo{author}{\bibfnamefont{S.}~\bibnamefont{Yan}},
  \bibinfo{author}{\bibfnamefont{D.~A.} \bibnamefont{Huse}}, \bibnamefont{and}
  \bibinfo{author}{\bibfnamefont{S.~R.} \bibnamefont{White}},
  \bibinfo{journal}{Science} \textbf{\bibinfo{volume}{332}},
  \bibinfo{pages}{1173} (\bibinfo{year}{2011}).

\bibitem[{\citenamefont{Jiang et~al.}(2012{\natexlab{a}})\citenamefont{Jiang,
  Yao, and Balents}}]{jiang1}
\bibinfo{author}{\bibfnamefont{H.-C.} \bibnamefont{Jiang}},
  \bibinfo{author}{\bibfnamefont{H.}~\bibnamefont{Yao}}, \bibnamefont{and}
  \bibinfo{author}{\bibfnamefont{L.}~\bibnamefont{Balents}},
  \bibinfo{journal}{Phys. Rev. B} \textbf{\bibinfo{volume}{86}},
  \bibinfo{pages}{024424} (\bibinfo{year}{2012}{\natexlab{a}}).

\bibitem[{\citenamefont{Jiang et~al.}(2012{\natexlab{b}})\citenamefont{Jiang,
  Wang, and Balents}}]{jiang2}
\bibinfo{author}{\bibfnamefont{H.-C.} \bibnamefont{Jiang}},
  \bibinfo{author}{\bibfnamefont{Z.}~\bibnamefont{Wang}}, \bibnamefont{and}
  \bibinfo{author}{\bibfnamefont{L.}~\bibnamefont{Balents}},
  \bibinfo{journal}{Nature Physics} \textbf{\bibinfo{volume}{8}},
  \bibinfo{pages}{902} (\bibinfo{year}{2012}{\natexlab{b}}).

\bibitem[{\citenamefont{Isakov et~al.}(2011)\citenamefont{Isakov, Hastings, and
  Melko}}]{melko1}
\bibinfo{author}{\bibfnamefont{S.~V.} \bibnamefont{Isakov}},
  \bibinfo{author}{\bibfnamefont{M.~B.} \bibnamefont{Hastings}},
  \bibnamefont{and} \bibinfo{author}{\bibfnamefont{R.~G.} \bibnamefont{Melko}},
  \bibinfo{journal}{Nature Physics} \textbf{\bibinfo{volume}{7}},
  \bibinfo{pages}{772} (\bibinfo{year}{2011}).

\bibitem[{\citenamefont{Isakov et~al.}(2012)\citenamefont{Isakov, Hastings, and
  Melko}}]{melko2}
\bibinfo{author}{\bibfnamefont{S.~V.} \bibnamefont{Isakov}},
  \bibinfo{author}{\bibfnamefont{M.~B.} \bibnamefont{Hastings}},
  \bibnamefont{and} \bibinfo{author}{\bibfnamefont{R.~G.} \bibnamefont{Melko}},
  \bibinfo{journal}{Science} \textbf{\bibinfo{volume}{335}},
  \bibinfo{pages}{193} (\bibinfo{year}{2012}).

\bibitem[{\citenamefont{Shimizu et~al.}(2003)\citenamefont{Shimizu, Miyagawa,
  Kanoda, Maesato, and Saito}}]{Shimuzi03_PRL_91_107001}
\bibinfo{author}{\bibfnamefont{Y.}~\bibnamefont{Shimizu}},
  \bibinfo{author}{\bibfnamefont{K.}~\bibnamefont{Miyagawa}},
  \bibinfo{author}{\bibfnamefont{K.}~\bibnamefont{Kanoda}},
  \bibinfo{author}{\bibfnamefont{M.}~\bibnamefont{Maesato}}, \bibnamefont{and}
  \bibinfo{author}{\bibfnamefont{G.}~\bibnamefont{Saito}},
  \bibinfo{journal}{Phys. Rev. Lett.} \textbf{\bibinfo{volume}{91}},
  \bibinfo{pages}{107001} (\bibinfo{year}{2003}).

\bibitem[{\citenamefont{Kurosaki et~al.}(2005)\citenamefont{Kurosaki, Shimizu,
  Miyagawa, Kanoda, and Saito}}]{Kurosaki05_PRL_95_177001}
\bibinfo{author}{\bibfnamefont{Y.}~\bibnamefont{Kurosaki}},
  \bibinfo{author}{\bibfnamefont{Y.}~\bibnamefont{Shimizu}},
  \bibinfo{author}{\bibfnamefont{K.}~\bibnamefont{Miyagawa}},
  \bibinfo{author}{\bibfnamefont{K.}~\bibnamefont{Kanoda}}, \bibnamefont{and}
  \bibinfo{author}{\bibfnamefont{G.}~\bibnamefont{Saito}},
  \bibinfo{journal}{Phys. Rev. Lett.} \textbf{\bibinfo{volume}{95}},
  \bibinfo{pages}{177001} (\bibinfo{year}{2005}).

\bibitem[{\citenamefont{Itou et~al.}(2009)\citenamefont{Itou, Oyamada, Maegawa,
  Tamura, and Kato}}]{131dmit1}
\bibinfo{author}{\bibfnamefont{T.}~\bibnamefont{Itou}},
  \bibinfo{author}{\bibfnamefont{A.}~\bibnamefont{Oyamada}},
  \bibinfo{author}{\bibfnamefont{S.}~\bibnamefont{Maegawa}},
  \bibinfo{author}{\bibfnamefont{M.}~\bibnamefont{Tamura}}, \bibnamefont{and}
  \bibinfo{author}{\bibfnamefont{R.}~\bibnamefont{Kato}},
  \bibinfo{journal}{Journal of Physics: Conference Series}
  \textbf{\bibinfo{volume}{145}}, \bibinfo{pages}{012039}
  (\bibinfo{year}{2009}).

\bibitem[{\citenamefont{Itou et~al.}(2008)\citenamefont{Itou, Oyamada, Maegawa,
  Tamura, and Kato}}]{131dmit2}
\bibinfo{author}{\bibfnamefont{T.}~\bibnamefont{Itou}},
  \bibinfo{author}{\bibfnamefont{A.}~\bibnamefont{Oyamada}},
  \bibinfo{author}{\bibfnamefont{S.}~\bibnamefont{Maegawa}},
  \bibinfo{author}{\bibfnamefont{M.}~\bibnamefont{Tamura}}, \bibnamefont{and}
  \bibinfo{author}{\bibfnamefont{R.}~\bibnamefont{Kato}},
  \bibinfo{journal}{Physical Review B} \textbf{\bibinfo{volume}{77}},
  \bibinfo{pages}{104413} (\bibinfo{year}{2008}).

\bibitem[{\citenamefont{Shimizu et~al.}(2007)\citenamefont{Shimizu, Akimoto,
  Tsujii, Tajima, and Kato}}]{131dmit3}
\bibinfo{author}{\bibfnamefont{Y.}~\bibnamefont{Shimizu}},
  \bibinfo{author}{\bibfnamefont{H.}~\bibnamefont{Akimoto}},
  \bibinfo{author}{\bibfnamefont{H.}~\bibnamefont{Tsujii}},
  \bibinfo{author}{\bibfnamefont{A.}~\bibnamefont{Tajima}}, \bibnamefont{and}
  \bibinfo{author}{\bibfnamefont{R.}~\bibnamefont{Kato}},
  \bibinfo{journal}{Journal of Physics: Condensed Matter}
  \textbf{\bibinfo{volume}{19}}, \bibinfo{pages}{145240}
  (\bibinfo{year}{2007}).

\bibitem[{\citenamefont{Grover et~al.}(2010)\citenamefont{Grover, Trivedi,
  Senthil, and Lee}}]{Grover10_PRB_81_245121}
\bibinfo{author}{\bibfnamefont{T.}~\bibnamefont{Grover}},
  \bibinfo{author}{\bibfnamefont{N.}~\bibnamefont{Trivedi}},
  \bibinfo{author}{\bibfnamefont{T.}~\bibnamefont{Senthil}}, \bibnamefont{and}
  \bibinfo{author}{\bibfnamefont{P.~A.} \bibnamefont{Lee}},
  \bibinfo{journal}{Phys. Rev. B} \textbf{\bibinfo{volume}{81}},
  \bibinfo{pages}{245121} (\bibinfo{year}{2010}).

\bibitem[{\citenamefont{Qi et~al.}(2009)\citenamefont{Qi, Xu, and
  Sachdev}}]{qixu}
\bibinfo{author}{\bibfnamefont{Y.}~\bibnamefont{Qi}},
  \bibinfo{author}{\bibfnamefont{C.}~\bibnamefont{Xu}}, \bibnamefont{and}
  \bibinfo{author}{\bibfnamefont{S.}~\bibnamefont{Sachdev}},
  \bibinfo{journal}{Phys. Rev. Lett.} \textbf{\bibinfo{volume}{102}},
  \bibinfo{pages}{176401} (\bibinfo{year}{2009}).

\bibitem[{\citenamefont{Biswas et~al.}(2011)\citenamefont{Biswas, Fu, Laumann,
  and Sachdev}}]{rudro}
\bibinfo{author}{\bibfnamefont{R.~R.} \bibnamefont{Biswas}},
  \bibinfo{author}{\bibfnamefont{L.}~\bibnamefont{Fu}},
  \bibinfo{author}{\bibfnamefont{C.~R.} \bibnamefont{Laumann}},
  \bibnamefont{and} \bibinfo{author}{\bibfnamefont{S.}~\bibnamefont{Sachdev}},
  \bibinfo{journal}{Phys. Rev. B} \textbf{\bibinfo{volume}{83}},
  \bibinfo{pages}{245131} (\bibinfo{year}{2011}).

\bibitem[{\citenamefont{Barkeshli et~al.}(2013)\citenamefont{Barkeshli, Yao,
  and Kivelson}}]{kivelson}
\bibinfo{author}{\bibfnamefont{M.}~\bibnamefont{Barkeshli}},
  \bibinfo{author}{\bibfnamefont{H.}~\bibnamefont{Yao}}, \bibnamefont{and}
  \bibinfo{author}{\bibfnamefont{S.~A.} \bibnamefont{Kivelson}},
  \bibinfo{journal}{Phys. Rev. B} \textbf{\bibinfo{volume}{87}},
  \bibinfo{pages}{140402(R)} (\bibinfo{year}{2013}).

\bibitem[{\citenamefont{Mishmash et~al.}(2013)\citenamefont{Mishmash, Garrison,
  Bieri, and Xu}}]{xudid}
\bibinfo{author}{\bibfnamefont{R.~V.} \bibnamefont{Mishmash}},
  \bibinfo{author}{\bibfnamefont{J.~R.} \bibnamefont{Garrison}},
  \bibinfo{author}{\bibfnamefont{S.}~\bibnamefont{Bieri}}, \bibnamefont{and}
  \bibinfo{author}{\bibfnamefont{C.}~\bibnamefont{Xu}},
  \bibinfo{journal}{arXiv:1307.0829}  (\bibinfo{year}{2013}).

\bibitem[{\citenamefont{Yao et~al.}(2010)\citenamefont{Yao, Fu, and
  Qi}}]{yaoqi}
\bibinfo{author}{\bibfnamefont{H.}~\bibnamefont{Yao}},
  \bibinfo{author}{\bibfnamefont{L.}~\bibnamefont{Fu}}, \bibnamefont{and}
  \bibinfo{author}{\bibfnamefont{X.-L.} \bibnamefont{Qi}},
  \bibinfo{journal}{arXiv:1012.4470}  (\bibinfo{year}{2010}).

\bibitem[{\citenamefont{Lu and Vishwanath}(2013)}]{luset}
\bibinfo{author}{\bibfnamefont{Y.-M.} \bibnamefont{Lu}} \bibnamefont{and}
  \bibinfo{author}{\bibfnamefont{A.}~\bibnamefont{Vishwanath}},
  \bibinfo{journal}{arXiv:1302.2634}  (\bibinfo{year}{2013}).

\bibitem[{\citenamefont{Kitaev}(2003)}]{kitaev2003}
\bibinfo{author}{\bibfnamefont{A.~Y.} \bibnamefont{Kitaev}},
  \bibinfo{journal}{Ann. Phys.} \textbf{\bibinfo{volume}{303}},
  \bibinfo{pages}{1} (\bibinfo{year}{2003}).

\bibitem[{\citenamefont{Grover and Senthil}(2008)}]{groversenthil}
\bibinfo{author}{\bibfnamefont{T.}~\bibnamefont{Grover}} \bibnamefont{and}
  \bibinfo{author}{\bibfnamefont{T.}~\bibnamefont{Senthil}},
  \bibinfo{journal}{Phys. Rev. Lett.} \textbf{\bibinfo{volume}{100}},
  \bibinfo{pages}{156804} (\bibinfo{year}{2008}).

\bibitem[{\citenamefont{Schnyder et~al.}(2009)\citenamefont{Schnyder, Ryu,
  Furusaki, and Ludwig}}]{ludwigclass1}
\bibinfo{author}{\bibfnamefont{A.~P.} \bibnamefont{Schnyder}},
  \bibinfo{author}{\bibfnamefont{S.}~\bibnamefont{Ryu}},
  \bibinfo{author}{\bibfnamefont{A.}~\bibnamefont{Furusaki}}, \bibnamefont{and}
  \bibinfo{author}{\bibfnamefont{A.~W.~W.} \bibnamefont{Ludwig}},
  \bibinfo{journal}{AIP Conf. Proc.} \textbf{\bibinfo{volume}{1134}},
  \bibinfo{pages}{10} (\bibinfo{year}{2009}).

\bibitem[{\citenamefont{Ryu et~al.}(2010)\citenamefont{Ryu, Schnyder, Furusaki,
  and Ludwig}}]{ludwigclass2}
\bibinfo{author}{\bibfnamefont{S.}~\bibnamefont{Ryu}},
  \bibinfo{author}{\bibfnamefont{A.}~\bibnamefont{Schnyder}},
  \bibinfo{author}{\bibfnamefont{A.}~\bibnamefont{Furusaki}}, \bibnamefont{and}
  \bibinfo{author}{\bibfnamefont{A.}~\bibnamefont{Ludwig}},
  \bibinfo{journal}{New J. Phys.} \textbf{\bibinfo{volume}{12}},
  \bibinfo{pages}{065010} (\bibinfo{year}{2010}).

\bibitem[{\citenamefont{Yu}(2009)}]{kitaevclass}
\bibinfo{author}{\bibfnamefont{K.~A.} \bibnamefont{Yu}}, \bibinfo{journal}{AIP
  Conf. Proc} \textbf{\bibinfo{volume}{1134}}, \bibinfo{pages}{22}
  (\bibinfo{year}{2009}).

\bibitem[{\citenamefont{Fidkowski and Kitaev}(2010)}]{fidkowski1}
\bibinfo{author}{\bibfnamefont{L.}~\bibnamefont{Fidkowski}} \bibnamefont{and}
  \bibinfo{author}{\bibfnamefont{A.}~\bibnamefont{Kitaev}},
  \bibinfo{journal}{Phys. Rev. B} \textbf{\bibinfo{volume}{81}},
  \bibinfo{pages}{134509} (\bibinfo{year}{2010}).

\bibitem[{\citenamefont{Fidkowski and Kitaev}(2011)}]{fidkowski2}
\bibinfo{author}{\bibfnamefont{L.}~\bibnamefont{Fidkowski}} \bibnamefont{and}
  \bibinfo{author}{\bibfnamefont{A.}~\bibnamefont{Kitaev}},
  \bibinfo{journal}{Phys. Rev. B} \textbf{\bibinfo{volume}{83}},
  \bibinfo{pages}{075103} (\bibinfo{year}{2011}).

\bibitem[{\citenamefont{Qi}(2013)}]{qiz8}
\bibinfo{author}{\bibfnamefont{X.-L.} \bibnamefont{Qi}}, \bibinfo{journal}{New
  J. Phys.} \textbf{\bibinfo{volume}{15}}, \bibinfo{pages}{065002}
  (\bibinfo{year}{2013}).

\bibitem[{\citenamefont{Ryu and Zhang}(2012)}]{zhangz8}
\bibinfo{author}{\bibfnamefont{S.}~\bibnamefont{Ryu}} \bibnamefont{and}
  \bibinfo{author}{\bibfnamefont{S.-C.} \bibnamefont{Zhang}},
  \bibinfo{journal}{Phys. Rev. B} \textbf{\bibinfo{volume}{85}},
  \bibinfo{pages}{245132} (\bibinfo{year}{2012}).

\bibitem[{\citenamefont{Fidkowski et~al.}(2013)\citenamefont{Fidkowski, Chen,
  and Vishwanath}}]{chenhe3B}
\bibinfo{author}{\bibfnamefont{L.}~\bibnamefont{Fidkowski}},
  \bibinfo{author}{\bibfnamefont{X.}~\bibnamefont{Chen}}, \bibnamefont{and}
  \bibinfo{author}{\bibfnamefont{A.}~\bibnamefont{Vishwanath}},
  \bibinfo{journal}{Phys. Rev. X} \textbf{\bibinfo{volume}{3}},
  \bibinfo{pages}{041016} (\bibinfo{year}{2013}).

\bibitem[{\citenamefont{Wang and Senthil}(2014)}]{senthilhe3}
\bibinfo{author}{\bibfnamefont{C.}~\bibnamefont{Wang}} \bibnamefont{and}
  \bibinfo{author}{\bibfnamefont{T.}~\bibnamefont{Senthil}},
  \bibinfo{journal}{Phys. Rev. B} \textbf{\bibinfo{volume}{89}},
  \bibinfo{pages}{195124} (\bibinfo{year}{2014}).

\bibitem[{\citenamefont{You and Xu}(2014)}]{youinversion}
\bibinfo{author}{\bibfnamefont{Y.-Z.} \bibnamefont{You}} \bibnamefont{and}
  \bibinfo{author}{\bibfnamefont{C.}~\bibnamefont{Xu}}, \bibinfo{journal}{Phys.
  Rev. B} \textbf{\bibinfo{volume}{90}}, \bibinfo{pages}{245120}
  (\bibinfo{year}{2014}).

\bibitem[{\citenamefont{Metlitski et~al.}()\citenamefont{Metlitski, Fidkowski,
  Chen, and Vishwanath}}]{max14}
\bibinfo{author}{\bibfnamefont{M.~A.} \bibnamefont{Metlitski}},
  \bibinfo{author}{\bibfnamefont{L.}~\bibnamefont{Fidkowski}},
  \bibinfo{author}{\bibfnamefont{X.}~\bibnamefont{Chen}}, \bibnamefont{and}
  \bibinfo{author}{\bibfnamefont{A.}~\bibnamefont{Vishwanath}},
  \eprint{arXiv:1406.3032}.

\end{thebibliography}

\end{document}